\documentstyle[epsf]{crnart11}
\newcommand{\beq}{\begin{equation}}
\newcommand{\eeq}{\end{equation}}

\begin{document}
\begin{Ntitlepage}
\docnum{CERN--TH-6956/93}
\docnum{FSU-SCRI-93-87}
\vspace{1cm}

\title{\Large{On Spin and Matrix Models in the Complex Plane}}
\vspace*{1cm}

\author{{\sc Poul H. Damgaard} \\
CERN -- Geneva \\
{}~\\
{\sc Urs M. Heller} \\
Supercomputer Computations Research Institute \\
The Florida State University \\
Tallahassee, FL 32306}

\begin{abstract}
We describe various aspects of statistical mechanics defined in the
complex temperature or coupling-constant plane. Using exactly
solvable models, we analyse such aspects as renormalization group
flows in the complex plane, the distribution of partition function
zeros, and the question of new coupling-constant symmetries of
complex-plane spin models. The double-scaling form of matrix
models is shown to be exactly equivalent to finite-size scaling
of 2-dimensional spin systems. This is used to show that the
string susceptibility exponents derived from matrix models can be
obtained numerically with very high accuracy from the scaling of
finite-$N$ partition function zeros in the complex plane.
\end{abstract}

\vspace{40mm}
\begin{flushleft}
CERN--TH-6956/93 \\
FSU-SCRI-93-87 \\
July 1993 \\
\end{flushleft}
\end{Ntitlepage}
\newpage


\section{Introduction} \label{Intro}

In contrast with what might be expected at first sight, the generalization
of spin models and gauge theories to the complex activity \cite{Lee} or
temperature \cite{Fisher} plane is a subject rich in physical content.
In general, information concerning the phase structure, and, in particular,
the possible phase transitions for physical values of the parameters can
be extracted from the behaviour of the finite volume partition function
in the complex plane. The Yang-Lee
edge singularity \cite{Lee,Fisher1} is one prime example of this, the
predictions of the distribution of partition function zeros in the
complex plane another \cite{Itzykson}. That these beautiful results can
also be very much of practical use in the study of phase transitions, has
been amply demonstrated \cite{Marinari}.

The purpose of this paper is to consider some aspects of spin, matrix
and gauge
theories in the complex plane that may have gone unnoticed until now. In so
far as it is possible, we shall compare general predictions for the
complex plane behaviour with exact analytical solutions. By doing so, we
unravel various subtleties involved in defining these models for complex
parameters, and we also gain additional insight into the nature of
statistical mechanics off the real axis.

The simplest example of an exactly solvable model in statistical
mechanics is the one-dimensional Ising model. Although this model cannot
have a real phase transition at non-zero temperature, it does in fact have
a continuous phase transition at $T = 0$. Because of the special nature of
a phase transition occurring at the origin, not all of the formalism of
critical phenomena can directly be applied to this system. In particular,
only certain {\em ratios} of critical exponents are uniquely defined.
Nevertheless, this is often sufficient to test the more general predictions
of complex plane partition functions. In section~\ref{1DIsing}, we use this
model to illustrate some of these more general results concerning the
location of partition function zeros \cite{Itzykson}, and use it to comment
on some interesting aspects of renormalization group flows in the complex
plane. This is possible, since one can construct {\em exact}
renormalization group transformations in the space of just three operators
(the energy, the spin and the constant operator) by means of spin
decimation.

In section~\ref{NewSym} we turn to the question of new symmetries of
discrete spin models at complex temperature, an issue recently brought up
in the case of the Ising model by Marchesini and Shrock \cite{Shrock}. We
show here that  the non-trivial part of these new symmetries requires a
careful definition of what is meant by the thermodynamic limit in the
complex plane.

Section~\ref{DoubleScal} contains a discussion of the double scaling limit
in matrix models and a discussion of 2D $U(N)$ lattice gauge theory in the
complex coupling constant plane. The parameter $N$, which eventually is
sent to infinity in order to obtain a phase transition \cite{Witten}, is
shown to play a r\^{o}le quite analogous to the volume in ordinary systems
with a finite number of degrees of freedom per site (or plaquette).  In
this manner, we demonstrate that the double scaling of matrix models (and
hence  low-dimensional string theory) is nothing but ordinary
{\em finite-size scaling} of models in statistical mechanics. This allows
us to extract
numerically the ``string susceptibility" exponent of matrix models through
the scaling of their corresponding partition function zeros. Finally,
section~\ref{Conclusions} contains our conclusions.

\section{The 1D Ising Model at Complex Activity and Complex Temperature}
\label{1DIsing}

Define the partition function for a finite number of sites $N$ as
\beq
{\cal Z}_N ~=~ \sum_{\sigma} {\rm e}^{K\sum_{j=1}^N \sigma_j\sigma_{j+1}
+ h \sum_{j=1}^N \sigma_j} ~,
\eeq
with periodic boundary conditions, i.e.,$ ~\sigma_{N+1} = \sigma_1$.
Introducing the symmetric transfer matrix $V(\sigma,\sigma')$, it is
trivial to compute the partition function from the trace:
\beq
{\cal Z}_N ~=~ {\rm Tr} V^N ~=~ \lambda_+^N + \lambda_-^N,
\label{eq:Z1D}
\eeq
where $\lambda_+$ and $\lambda_-$ are the two eigenvalues. Noting that
$V(\sigma,\sigma')$ can be represented by a $2\times 2$ matrix,
\begin{eqnarray}
V(\sigma, \sigma') &= & {\rm e}^{ K \sigma \sigma' +
 \frac{1}{2} (\sigma + \sigma') } \cr
& = & \left( \begin{array}{cc} {\rm e}^{K+h} & {\rm e}^{-K} \\
                               {\rm e}^{-K}  & {\rm e}^{K-h}
      \end{array} \right)
\end{eqnarray}
one has
\beq
\lambda_{\pm} = {\rm e}^K \cosh(h) \pm \sqrt{{\rm e}^{2K}\sinh^2(h) +
 {\rm e}^{-2K}}.
\label{eq:lpm}
\eeq
This is all well known. On the real axis $\lambda_+ > \lambda_-$, and the
thermodynamic limit of the partition function,
reached as $N \to \infty$, involves only $\lambda_+$.
For example, the free energy per spin is found to be
\beq
f = \lim_{N \to \infty}N^{-1} \ln {\cal Z}_N = \ln \lambda_+.
\label{eq:FE_infV}
\eeq
Similarly, the magnetization per spin is explicitly given by
\begin{eqnarray}
m(h,K) & ~=~ & \lim_{N \to \infty} N^{-1} {\cal Z}_N^{-1}
 \frac{\partial}{\partial h}\left[ {\cal Z}_N \right] \nonumber \\
& ~=~ & \frac{ {\rm e}^K\sinh(h)}{[{\rm e}^{2K}\sinh^2(h) +
 {\rm e}^{-2K}]^{1/2}}. \label{eq:mag1D}
\end{eqnarray}

It is also well-known that this system cannot have a phase transition for
non-zero temperatures, and any real value of the magnetic field.  We see
this explicitly from the solution (\ref{eq:FE_infV}). Nevertheless, at $h =
0$ the system {\em does} have a continuous phase transition at $T = 0 ~(K =
\infty)$, which to a very large extent can be phrased within the standard
formalism of critical phenomena \cite{Nelson}. Since the transition occurs
at an asymmetric point at which the usual scaled temperature variable $t =
(T - T_c)/T_c$ cannot be defined, one must be more careful and  rewrite all
scaling relations in terms of the correlation length. This in turn  is
computed from the 2-point function \cite{Baxter},
\beq
g_{ij} ~\equiv~ \langle\sigma_i\sigma_j\rangle - \langle\sigma_i\rangle
\langle\sigma_j\rangle ~=~
\sin^2(2\phi)[\lambda_+/\lambda_-]^{|i-j|},
\eeq
where $\phi$ is determined from the solution $0 < \phi < \pi/2$ of
$\cot(2\phi) = {\rm e}^{2K}\sinh(h)$. This shows that
\beq
\xi = 1/\ln[\lambda_+/\lambda_-],
\eeq
which at $h = 0$ is well-defined (and non-singular) for all positive
temperatures. The usual critical exponents are then defined through the
behaviour of the singular parts of the thermodynamic quantities as one
approaches the critical point at $T = 0$:
\beq
\Delta f(0,T) \sim \xi^{-(2-\alpha)/\nu}~,~~~ m(0,T) \sim \xi^{-\beta/\nu}~,~~
\chi(0,T) \sim \xi^{\gamma/\nu}. \label{eq:critexp}
\eeq

{}From the exact solution (\ref{eq:FE_infV}) it follows that
\beq
\gamma = \nu = 2 - \alpha,
\eeq
while from the corresponding exact solution of the 2-point function we see
that $\eta = 1$. Finally, from the behaviour of the magnetization at
non-vanishing magnetic  field {\em at} the critical point, $m(h,0) =
sgn(h)$, it follows that $\delta = \infty$. Although this model does not
display spontaneous  magnetization, it follows from the definition
(\ref{eq:critexp}) that $\beta = 0$. These values of the critical exponents
are all consistent with the general scaling relations
\begin{eqnarray}
\alpha + 2 \beta + \gamma & = & 2 \cr
\beta\delta & = & \beta + \gamma \cr
\gamma & = & (2 - \eta)\nu
\end{eqnarray}
for $\beta\delta = \nu$ (a result which also follows formally from the exact
solution given above), and with the general hyperscaling relation
\begin{eqnarray}
d\nu & = & 2 - \alpha \cr
\delta & = & (d + 2 - \eta)/(d - 2 + \eta) \label{eq:hyperscal}
\end{eqnarray}
for $d = 1$. Thus, although not all critical exponents are uniquely defined
in this case, certain {\em ratios} of exponents are completely well defined,
and in full accord with general scaling relations at a critical point.

These simple results do not hold in general, however, once we go off the
real axis in either magnetic field $h$ (or activity ${\rm e}^{-2h}$) or
temperature (proportional to $K^{-1}$). The reason is, of course, that the
two  transfer matrix eigenvalues $\lambda_{\pm}$ become in general
complex-valued. The simple thermodynamic limit $N \to \infty$ then
obviously does not always exist (if, $e.g.$, $|\lambda_{\pm}| = 1$, the
naive thermodynamic limit of the partition function is given by a
non-convergent phase running around on the unit circle). But there are
clearly also many regions in the complex plane where this naive
thermodynamic limit {\em is} well-defined. The location of partition
function zeros ${\cal Z}_N = 0$  is one such example, which we turn to
next.

One of the predictions of ref. \cite{Itzykson} is that for a finite system
of volume (number of sites) $N$, the partition function zeros close to
the critical point should scale as
\beq
h_n^2 ~=~ N^{-2\beta\delta/\nu} f_n(0),
\eeq
where the index $n$ indicates the $n$th partition function zero. Also,
the scaling function $f_n(KN^{1/\nu})$ evaluated at zero argument has
been predicted to behave, for $n$ large, like \cite{Itzykson}
\beq
f_n(0) ~=~ - C n^{2\delta/(\delta + 1)},
\eeq
where $C$ is some constant. In other words, close to the critical point
we should have
\beq
h_n^2 ~=~ - C N^{-2\beta\delta/\nu} n^{2\delta/(\delta + 1)},
\eeq
which in the case of the 1D Ising model unambiguously implies
\beq
h_n ~=~ i C^{1/2} \frac{n}{N}.
\label{eq:zeros1D}
\eeq

Let us compare this with the known exact solution. In the limit
$K >> 1$, it follows from eq.~(\ref{eq:lpm}) that
\begin{eqnarray}
\lambda_+ & ~=~ & {\rm e}^K \cosh(h) + {\rm e}^K \sinh(h) \cr
\lambda_- & ~=~ & {\rm e}^K \cosh(h) - {\rm e}^K \sinh(h).
\end{eqnarray}
Thus, in this limit we see from eq.~(\ref{eq:Z1D}) that
\beq
{\cal Z}_N ~\sim~ 2 {\rm e}^{NK} \cosh(Nh).
\eeq
The partition function zeros are therefore to be found at
\beq
h_n ~=~ \frac{i\pi}{2N} (2n + 1), \label{eq:zeros1Dn}
\eeq
or, since $n$ should be taken large in order to compare with eq.
(\ref{eq:zeros1D}),
\beq
h_n ~\sim~ i\pi \frac{n}{N},
\eeq
which is precisely the expected behaviour (\ref{eq:zeros1D}),
with $C = \pi^2$.

In the 2D Ising model the partition function zeros at zero magnetic field
lie on two intersecting circles in the complex temperature plane
\cite{Fisher}. One of these circles crosses the real temperature axis at
the 2D Ising critical temperature $T_c$ at a right angle. For the 1D Ising
model it follows from eq.~(\ref{eq:Z1D}) that the zero-field partition
function zeros lie on just one circle, the unit circle in the complex
$\tanh(K)$-plane, crossing the real axis at $\tanh(K) = 1$ (corresponding
to the 1D Ising critical coupling $K = \infty$) again at a right angle.
Although not rigorously related to
the general prediction for this angle $\varphi$ of ref.
\cite{Itzykson}:
\beq
\tan[(2-\alpha)\varphi] ~=~ \frac{\cos(\pi\alpha) - A_-/A_+}
{\sin(\pi\alpha)}, \label{eq:zero_angle}
\eeq
this behaviour is at least consistent with it. Here
$\alpha$ is the usual critical exponent, and $A_{\pm}$ are the
specific heat amplitudes above and below $T_c$. For $\varphi = \pi/2$
eq.~(\ref{eq:zero_angle}) can be brought to
\beq
[ 1  - A_-/A_+ ] \cos(\alpha\pi) = - [ 1  - A_-/A_+ ] .
\eeq
Solutions are either $\alpha = 1$ with $A_-/A_+$ undetermined or $A_-/A_+ =
1$ with $\alpha$ undetermined. The latter agrees with what we previously
found and hence the prediction (\ref{eq:zero_angle}) applies  to the 1D
Ising model.

\subsection{\sc Complex-plane Singularities of the 1D Ising Model}

Let us define a `` critical point" anywhere in the complex plane
by the condition of a diverging correlation length, $\xi = \infty$.
In the 1D Ising model we have already seen that $\xi = 1/\ln[
\lambda_+/\lambda_-]$, where $\lambda_{\pm}$ are the two transfer
matrix eigenvalues. The condition $\xi = \infty$ in this case
implies $\lambda_+ = \lambda_-$. When $h = 0$, this gives
\beq
\cosh(K) ~=~ \sinh(K),
\eeq
which for real $K$ can only be satisfied asymptotically at $K = \infty$.
This is just the usual 1D Ising transition at $T = 0$.

\begin{figure}[htb]
\begin{center}
\leavevmode
\epsfxsize=280pt
\epsfbox[150 260 520 590]{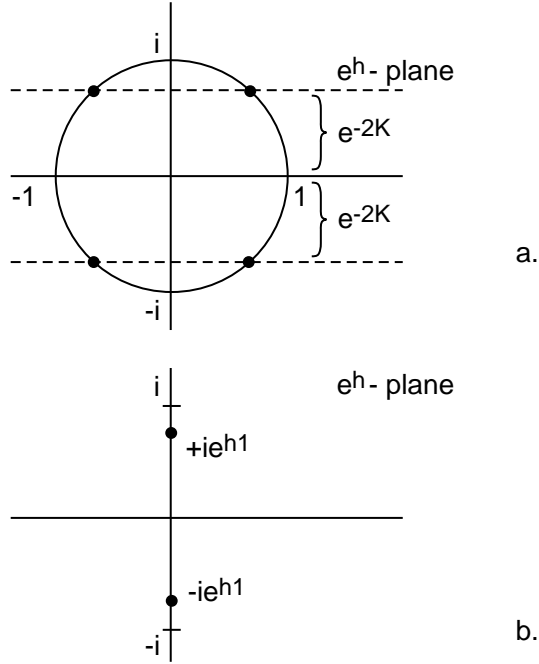}
\caption{\it Critical points for the 1D Ising model with complex couplings.}
\label{fig:Crit_comp}
\end{center}
\end{figure}

When $h \neq 0$, there are more possibilities. In particular, $\lambda_+
= \lambda_-$ in the whole subspace where
\beq
{\rm e}^{2K}\sinh^2(h) + {\rm e}^{-2K} ~=~ 0. \label{eq:crit_cond}
\eeq
Consider first a simple special case, where $K$ is real, but $h$ is allowed
to become complex. Then, parametrizing $h = h_1 + ih_2$, and using some
trigonometric identities, we find that vanishing of the imaginary part of
eq.~(\ref{eq:crit_cond}) implies
\beq
2 {\rm e}^{2K}\sinh(h_1)\cos(h_2)\cosh(h_1)\sin(h_2) ~=~ 0,
\eeq
$i.e.,~ h_1 = 0$, or $h_2 = n\pi/2$ or {\em both}. Demanding also the
vanishing of the real part of eq.~(\ref{eq:crit_cond}) thus splits up in
two parts: (a) $h_1 =0$ and (b) $h_2 = n\pi/2$. Case (a) implies
\beq
\sin^2(h_2) ~=~ \exp[-4K],
\eeq
which can only be satisfied for $K > 0$. The solution to this equation in
the complex ${\rm e}^h$-plane is shown graphically in
fig.~\ref{fig:Crit_comp}a.

Case (b) leads to
\beq
{\rm e}^{2K}\left\{\sinh^2(h_1)\cos^2\left(\frac{n\pi}{2}\right) -
\cosh^2(h_1) \left[1 - \cos^2\left(\frac{n\pi}{2}\right)\right]\right\} +
{\rm e}^{-2K} ~=~ 0,
\eeq
which in turn has two possible solutions. In the first case,
$\cos^2(n\pi/2) = +1$,
for which the above equation reduces to ${\rm e}^{2K}\sinh^2(h_1) +
{\rm e}^{-2K} = 0$. This cannot be satisfied for real $h_1$ and $K$.
The other case is $\cos^2(n\pi/2) = 0,~i.e.~ n = (2j+1), ~j = 0, 1, \ldots$,
which leads to
\beq
\cosh^2(h_1) ~=~ \exp[-4K].
\eeq
Although this equation cannot be satisfied for $K > 0$, there {\em are}
non-trivial solutions for $K < 0$, which we can think of as the
antiferromagnetic situation. In the complex ${\rm e}^h$-plane this corresponds
to the solutions shown in fig.~\ref{fig:Crit_comp}b.

At these points in the complex plane we have not only  $\xi = \infty$, but
also magnetization $m = \infty$. The finite volume
partition function in these cases equals
\beq
{\cal Z}_N ~=~ 2[{\rm e}^K\cosh(h)]^N,
\eeq
so that, on account of $\cosh(h_1+ih_2) = \cosh(h_1)\cos(h_2) +
i\sinh(h_1)\sin(h_2)$, the case (a) corresponds to ${\cal Z}_N$ {\em real}.
In contrast, in the antiferromagnetic case of (b), $\lambda_+ =
\lambda_-$ are purely {\em imaginary}. The simplest thermodynamic limit
$N \to \infty$ does not exist in this case.

On viewing the above results, one question that immediately arises is: how
can a {\em local} quantity such as the magnetization per spin
diverge\footnote{ The absolute value of the magnetization
(\ref{eq:mag1D}) is clearly bound by
unity for all real values of $K$ and $h$, in agreement with the notion that
we are making a statistical average over a number fluctuating between +1
and $-$1. The magnetization escapes this bound in the complex plane by being
the average over complex-valued Boltzmann factor ``probabilities".} at
these complex-plane critical points?

To answer this question in detail, let us try to regularize everything by
considering first a finite volume of $N$ spins. We still have
\beq
m(h,K) ~\equiv~ {\cal Z}_N^{-1} \frac{\partial}{\partial h}
\left\{\frac{{\cal Z}_N}{N}\right\},
\eeq
but now both transfer matrix eigenvalues must be kept in the expression
for the partition function, $i.e.~ {\cal Z}_N = \lambda_+^N + \lambda_-^N$.
Carrying out the differentiation, a convenient way to express the
finite-volume magnetization is as
\beq
m(h,K) ~=~ \frac{{\rm e}^K \sinh(h)}{[{\rm e}^{2K}\sinh^2(h) +
 {\rm e}^{-2K}]^{1/2}}
 ~\frac{\lambda_+^N - \lambda_-^N}{\lambda_+^N + \lambda_-^N}.
\eeq

Ordinarily, on the real axes, $\lambda_+ > \lambda_-$, in which case the
above expression reduces in the thermodynamic limit $N \to \infty$ to the
magnetization $m(h,K)$ of eq.~(\ref{eq:mag1D}). But the present case
corresponds to $\lambda_+ = \lambda_-$, so we must be more careful. If we
take the limit $\lambda_- \to \lambda_+$ (where ${\rm e}^{2K}\sinh^2(h) +
{\rm e}^{-2K} \to 0$) for {\em any finite} $N$, then
\beq
\lambda_+^N - \lambda_-^N ~\simeq~ [{\rm e}^K\cosh(h)]^{N-1}\times 2N \times
({\rm e}^{2K}\sinh^2(h) + {\rm e}^{-2K})^{1/2},
\eeq
while of course
\beq
\lambda_+^N + \lambda_-^N ~\simeq~ 2[{\rm e}^K\cosh(h)]^N
\eeq
to lowest order. Thus at any finite $N$ we have
\beq
m(h,K) = \frac{N {\rm e}^K\sinh(h)}{{\rm e}^K\cosh(h)} = N \tanh(h)
\label{eq:N}
\eeq
at these complex-plane critical points. This behaviour, linear growth with
the size of the system, is actually precisely
as expected from finite-size scaling theory when we treat these complex-plane
singularities as true critical points. To see this, consider again the
infinite-volume magnetization at, for simplicity, purely imaginary
external magnetic field $h = ih_2$,
\beq
m(h_2,K) = \frac{ie^K\sin(h_2)}{[e^{-2K} - \sin^2(h_2)e^{2K}]^{1/2}} ~.
\eeq
Expanding $h_2 = h_2^* + \Delta h_2$ around a critical magnetic field
$h_2^*$ given by $\sin^2(h_2^*) = \exp[-4K]$, we note that
\beq
m(\Delta h_2,K) ~\sim~ \Delta h_2^\rho
\eeq
as $\Delta h_2 \to 0$, with $\rho = -1/2$.

Similarly, for the correlation length we have
\beq
\xi^{-1} = \ln\left[\frac{\lambda_+}{\lambda_-}\right] =
\ln\left[\frac{e^K\cos(h_2^*) + c\sqrt{\Delta h_2}}{e^K\cos(h_2^*) -
c\sqrt{\Delta h_2}} \right] ~\sim~ \Delta h_2^\nu  ~,
\eeq
with $\nu = 1/2$. Finite-size scaling theory \cite{Barber} then predicts
\beq
m(\Delta h_2,K) ~\sim~ N^{-\rho/\nu} ~\sim~ N ~,
\eeq
in agreement with the explicit solution (\ref{eq:N}).

\subsection{\sc Renormalization Group Flows in the Complex Plane}

Not surprisingly, the 1D Ising model can be solved by an {\em exact}
real-space renormalization group transformation, spin decimation. This hence
gives us the possibility of explicitly observing some exact results
concerning RG flows in the complex plane, results that cannot easily be
obtained from more general principles. For a rescaling factor $b = 2$, the
exact RG transformation for the 1D Ising model takes the following form
\cite{Nelson}:
\begin{eqnarray}
\exp[4K'] &~=~& \frac{\cosh(2K + h)\cosh(2K - h)}{\cosh^2(h)} \nonumber \\
\exp[2h'] &~=~& {\rm e}^{2h} ~ \frac{\cosh(2K + h)}{\cosh(2K - h)},
\label{eq:RG1D}
\end{eqnarray}
where primed variables are the renormalized couplings. The RG flow in the
$K-h$ plane is not trivial \cite{Fisher1}, but when $h = 0$ it becomes
very simple: we get $\exp[2h'] = 1, ~i.e.~ h' = 0$. This just
expresses consistency; when we start with no magnetic field, we do not
generate it in the process of renormalization. The flow in the temperature
direction is slightly less trivial:
\beq
{\rm e}^{4K'} ~=~ \cosh^2(2K).
\eeq
This equation has two fixed points, $K^* =0$ and $K^* = \infty$. The fixed
point at $K^* = \infty$ is unstable, and it is the usual 1D Ising critical
point; here $\xi = \infty$.

Now, from general principles the infinite correlation length condition
$\xi = \infty$, which is fulfilled on the surface
\beq
\sinh^2(h) ~=~ - {\rm e}^{-4K}, \label{eq:xi_inf}
\eeq
must be associated with a ``critical hypersurface" in the complex plane.
Once on this surface, the RG flow must remain on it. It is easy to
verify that
\begin{eqnarray}
{\rm e}^{-4K'} &~=~& -\sinh^2(h') ~~~{\mbox{\rm if and only if}} \cr
{\rm e}^{-4K} &~=~& -\sinh^2(h).
\end{eqnarray}
Note that this  complex-plane hypersurface intersects the plane with real
couplings $K$ and $h$ at the usual critical point $(K,h) = (\infty,0)$, so
we should expect that this surface in almost all respects can be treated as
a standard critical hypersurface. (There should then also be complex-plane
equivalents of irrelevant operators, pushing the RG flow towards a fixed
point on this surface).

\begin{figure}[htb]
\begin{center}
\leavevmode
\epsfxsize=280pt
\epsfbox[20 30 620 500]{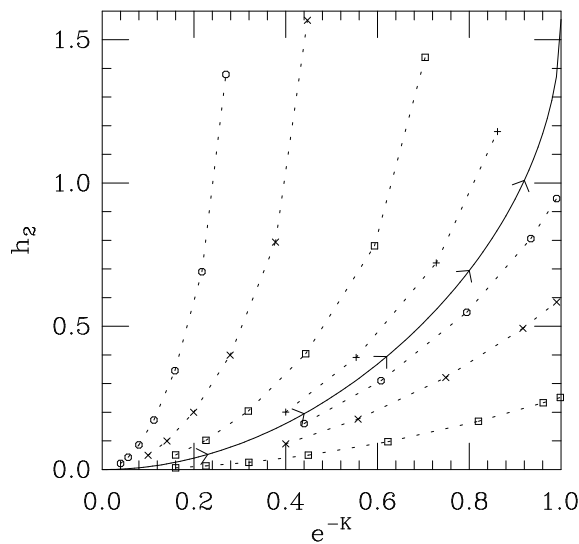}
\caption{\it Renormalization group flow for the 1D Ising model in the plane
${\rm e}^{-K}$ and $h=ih_2$. The critical line is shown with the direction
of flow indicated. Some results from iterating the RG away from the
critical line are also given with different plotting symbols and connected
with dotted lines to guide the eye. The flow is always away from the origin.}
\label{fig:RG_Ising}
\end{center}
\end{figure}

It follows from eqs.~(\ref{eq:RG1D}) that the critical surface given by
(\ref{eq:xi_inf}) contains, besides the unstable fixed point $(K,h) =
(\infty,0)$, two stable ones at $(K,h) = (0,\pm i \pi/2)$ (everything is
symmetric under $h \to -h$). The critical line in the interesting coupling
plane $({\rm e}^{-K},ih_2)$ and the RG flow on it are shown in
fig.~\ref{fig:RG_Ising}. The figure also shows the results of iterating the
RG equations starting at several points a little off of the critical line.

What about the {\em zeros} of ${\cal Z}_N$? Let us start with a lattice of
couplings $(K,h=0)$ and a number of sites equal to $2N$, and then apply the RG
transformation of rescaling factor 2. The zeros of ${\cal Z}_{2N}$ are given
by
\beq
\frac{{\rm e}^K - {\rm e}^{-K}}{{\rm e}^K + {\rm e}^{-K}} ~=~
 {\rm e}^{i\pi(2n+1)/2N},
\eeq
and squaring this, we get successively
\begin{eqnarray}
{\rm e}^{i\pi(2n+1)/N} &~=~& \left(\frac{{\rm e}^K - {\rm e}^{-K}}
{{\rm e}^K + {\rm e}^{-K}}\right)^2 \cr
&~=~& \frac{\cosh(2K) - 1}{\cosh(2K) + 1} \cr
&~=~& \frac{\cosh^{1/2}(2K) - \cosh^{-1/2}(2K)}
{\cosh^{1/2}(2K) + \cosh^{-1/2}(2K)} \cr
&~=~& \frac{{\rm e}^{K'} - {\rm e}^{-K'}}{{\rm e}^{K'} + {\rm e}^{-K'}}.
\end{eqnarray}
The zeros of the partition function thus behave as expected under this
exact RG transformation: the zeros of ${\cal Z}_{2N}$ move to those
of ${\cal Z}_N$.

The RG flow in the complex plane is therefore far from smooth, and except
for the neighbourhood of the critical hypersurface discussed above, it does
not much resemble the usual RG flows for real couplings. Consider plotting
the flow in the complex $\tanh(K)$-plane. On the unit circle of this plane
the partition function zeros jump around corresponding to going from
$\exp[i\pi(2n+1)/2N]$ to $\exp[i\pi(2n+1)/N]$ as $K$ goes into $K'$.
But in the ordinary coupling constant plane of $K$, this means going from,
for example in the last case $N = 1$:
$K = \pm i\pi/4$ to $K' = -\infty$ in {\em one} RG step.

\subsection{\sc The Generalized Heisenberg Model}

The partition function
\beq
{\cal{Z}}_N ~=~ \sum_{\vec{\sigma}} e^{K\sum_{j=1}^N \vec{\sigma}_j
\cdot \vec{\sigma}_{j+1} + \sum_{j=1}^N \vec{h}\cdot\vec{\sigma}_j}
\label{eq:Z_Heisen}
\eeq
defines a natural generalization of the 1D Ising model, -- the generalized
Heisenberg model in 1D. Here $\vec{\sigma}_i$ denotes an $n$-dimensional
vector of unit length, and $\vec{h}$ is a vector of fixed orientation
(which can be chosen at will) describing an external magnetic field. For
$n = 1$ this reduces to the ordinary 1D Ising model, while the cases
$n = 2$ and $n = 3$ correspond to the planar and original classical
Heisenberg model, respectively.
The model is solvable for all $n$ in one dimension when the external
field vanishes \cite{Stanley}. We shall here use this exact solution to
reinvestigate complex-temperature partition functions.

The eigenvalues of the transfer matrix of eq.~(\ref{eq:Z_Heisen}) are known
in closed analytical form \cite{Stanley,Balian}:
\beq
\lambda_i ~=~ (2\pi)^{\frac{1}{2}n} K^{1 - \frac{1}{2}n}
I_{\frac{1}{2}n-1+i}(K),
\eeq
where $i = 0,1, \ldots$, and $I_{\nu}(x)$ is the modified Bessel function
of order $\nu$. The model can be studied straightforwardly for integer
values of $n$, in which case it yields only small modifications in
comparison with the Ising case. (The phase transition temperature remains at
$T_c = 0$, the critical indices being equal to the Ising indices
for all finite values of $n$.) What is more interesting for our purpose
here is that the model also admits
a unique analytic continuation in $n$ for $0 < n < 1$ \cite{Balian}. In
this case the phase transition takes on a more familiar form, occurring
now at a finite value of $T_c$. The correlation function is determined
by the two leading transfer matrix eigenvalues, and reads explicitly
\beq
\langle \vec{\sigma}_i \vec{\sigma}_j \rangle ~=~
\left[\frac{I_{\frac{1}{2}n}(K)}{I_{\frac{1}{2}n-1}(K)}\right]^{|i-j|}.
\eeq
This corresponds to a correlation length
\beq
\xi^{-1} ~=~ \ln\left[\frac{I_{\frac{1}{2}n}(K)}{I_{\frac{1}{2}n-1}(K)}\right]
{}~,
\eeq
which diverges at the phase transition point $K_c$ determined by
\beq
I_{\frac{1}{2}n}(K_c) ~=~ I_{\frac{1}{2}n-1}(K_c).
\eeq
As mentioned above, this phase transition occurs at a finite value of
$K_c$ when $0 < n < 1$. The condition of a diverging correlation length,
\beq
|I_{\frac{1}{2}n}(K_c)| ~=~ |I_{\frac{1}{2}n-1}(K_c)|,
\eeq
continued into the complex plane thus reduces as $n \to 1$ to the circle
$|coth(K)| = 1$ of the 1D Ising model.

\begin{figure}[htb]
\begin{center}
\leavevmode
\epsfxsize=280pt
\epsfbox[20 30 620 500]{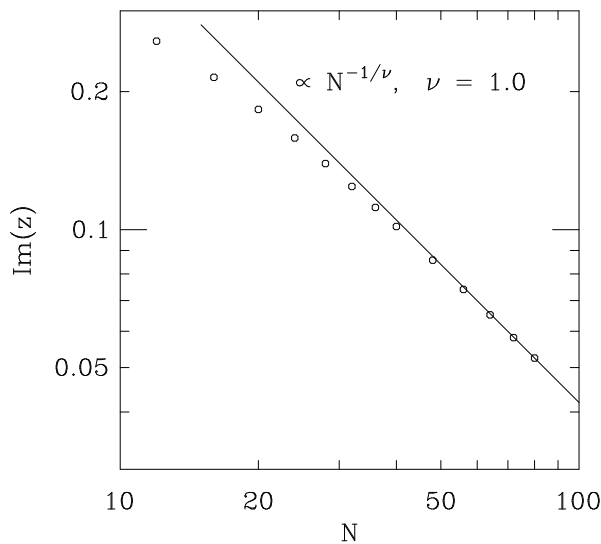}
\caption{\it Finite size scaling of the complex partition function zero
closest to the real axis for the $n = 0.5$ 1D Heisenberg model.}
\label{fig:Zero_scal_HB}
\end{center}
\end{figure}

Having solvable models with a finite critical point $K_c$, we can
investigate the finite-size scaling of the zeros of the partition function
in the complex plane \cite{Itzykson}. The zero closest to the real axis
should approach the critical point $K_c$ at a rate determined by the
correlation length exponent $\nu$
\beq
{\mbox{\rm Im}} (K_0) \propto N^{-1/\nu}~.   \label{eq:FSS_nu}
\eeq
As an example we consider $n = 0.5$ with the critical point at $K_c =
0.6755198$.  The scaling (\ref{eq:FSS_nu}) is shown in
fig.~\ref{fig:Zero_scal_HB}. From two successive sizes, we can get an
estimate $\nu(N)$ of $\nu$. These estimates are given in
table~\ref{tab:nu_zeros_HB}.

\begin{table}[hbt]
\begin{center}
\caption{\it Estimates of $\nu$ from the finite scaling of the imaginary
part of the partition function zeros computing from successive $N$ (second
column) and extrapolated to $N = \infty$ from two successive estimates
(third column).}
\label{tab:nu_zeros_HB}
\vskip 0.5cm
\begin{tabular}{|ccc|}
\hline
 $N$ & $\nu(N)$ & $\nu(\infty)$ \\
\hline
  8 &  3.168120 &  ~~ \\
 12 &  1.998528 &  1.062855 \\
 16 &  1.587125 &  1.058179 \\
 20 &  1.385652 &  1.027478 \\
 24 &  1.271576 &  1.012312 \\
 28 &  1.200939 &  1.005329 \\
 32 &  1.154338 &  1.002107 \\
 36 &  1.122063 &  1.000559 \\
 40 &  1.098848 &  0.999877 \\
 48 &  1.075312 &  1.021823 \\
 56 &  1.054356 &  0.996325 \\
 64 &  1.041037 &  0.997529 \\
 72 &  1.032060 &  0.998264 \\
 80 &  1.025729 &  0.998739 \\
\hline
\end{tabular}
\end{center}
\end{table}

We can see that the estimates for $\nu$ are already quite close to the
known exact value $\nu = 1$ \cite{Balian}. Assuming a finite-size
correction of the form $N^{-2}$, which fits the data best, we can get
infinite volume estimates from two successive estimates $\nu(N)$. These
estimates for $\nu$, also given in table~\ref{tab:nu_zeros_HB}, are
remarkably good.

\section{New Symmetries of Spin Models} \label{NewSym}

In a very interesting paper, Marchesini and Shrock \cite{Shrock}, have
recently derived what is apparently new symmetries of the
zero-field Ising model in any
number of dimensions. The symmetries are related to discrete transformations
of the Ising coupling constant (or temperature) {\em in the complex
temperature plane}. Evidence was found that these new symmetries are not
without influence on results derived for real coupling constants, since
they also signal new singularities in the complex plane, singularities
that may affect $e.g.$ Pad\'{e} approximants of
truncated series expansions on the real coupling constant axis.
The symmetries, although clearly not of universal nature, are
generalizable to other discrete spin models.
Again, let us consider these general results in the light of the simple
solvable 1D Ising model. We shall later make some related remarks on the
2D Ising case.

The 1D Ising model is of coordination number $q = 2$, and the two main
theorems of ref. \cite{Shrock} are therefore applicable in this case. These
state that the zero-field ($h = 0$) free energy $f(K)$
should be invariant under the following transformations:
\begin{eqnarray}
f(K) &~=~& f(K + in\pi)~, ~~~ n~{\mbox{\rm integer}} \label{eq:newSym1} \\
f(K) &~=~& f\left(K + i\left(n+\frac{1}{2}\right)\pi\right)~, ~~~
n~{\mbox{{\rm integer}}}.
\label{eq:newSym2}
\end{eqnarray}
But if we take the explicit expression for the  free energy
per spin of the 1D Ising model,
\beq
f(K) ~=~ - \ln[2\cosh(K)], \label{eq:freeE}
\eeq
this does not appear to be satisfied. In fact, the only invariance of this
kind in eq.~(\ref{eq:freeE}) is
\beq
K ~\to~ k + 2\pi i n ~,~~~ n~ {\mbox{\rm integer}},
\eeq
which is trivially satisfied by the Ising model in any number of
dimensions, on
account of the integer-valued nature of the spin-spin interaction. How
can we understand this discrepancy?

The explanation lies, of course, in the subtlety involved in defining the
thermodynamic limit of the partition function off the real axis. In
defining the free energy of the infinite volume Ising model as in
eq.~(\ref{eq:freeE}) above, we have already chosen one particular limit,
the one valid on the real coupling constant axis. Although this expression
can be analytically continued into the complex plane, the resulting free
energy may not be derivable from the infinite volume limit of the general
partition function in the complex plane. As we have already seen in several
examples before,  these two procedures do not always commute.

To understand the new symmetry, let us therefore revert to the finite
volume partition function ${\cal Z}_N = \lambda_+^N + \lambda_-^N$. With
zero magnetic field,
\beq
\lambda_+ ~=~ 2\cosh(K)~,~~~~\lambda_- ~=~ 2\sinh(K),
\eeq
and consider now the transformation (\ref{eq:newSym2}). Under this,
\begin{eqnarray}
\cosh(K) &~\to~& (-1)^n i \sinh(K) \cr
\sinh(K) &~\to~& (-1)^n i \cosh(K),
\end{eqnarray}
or in other words,
\beq
\lambda_+ ~\to (-1)^n i \lambda_- ~,~~~~ \lambda_- ~\to~ (-1)^n i \lambda_+.
\eeq

Consider now taking the thermodynamic limit in steps of $4N$ instead of
just in steps of $N$, as it is usually defined. Then, with $\tilde{N} =
4N$, the transformation (\ref{eq:newSym2}) acts on the partition function
as
\begin{eqnarray}
{\cal Z}_{\tilde{N}} &~=~& \lambda_+^{4N} + \lambda_-^{4N} \cr
 &~\to~& [(-1)^n i \lambda_-]^{4N} + [(-1)^n i \lambda_+]^{4N} \cr
&~=~& \lambda_-^{4N} + \lambda_+^{4N} ~=~ {\cal Z}_{\tilde{N}},
\end{eqnarray}
which shows explicitly that this indeed {\em is} a symmetry; the transformation
has not done more than what effectively amounts to swapping eigenvalues.

But here another subtlety in connection with this new symmetry becomes
apparent: it is not sufficient to remain at the level of finite volume
partition functions, perform the transformation, and then take  the
thermodynamic limit. Instead, {\em the symmetry hinges in an essential way
on the precise definition of boundary conditions and the very manner in
which this thermodynamic limit is defined}. On the real axis these problems
usually do not arise; one obtains a unique thermodynamic limit
independently of the manner in which the thermodynamic limit is
reached.\footnote{Special care is required in the case of spontaneous
symmetry breaking, even on the real temperature axis. This is perhaps
the closest we can get to an ordinary statistical mechanics analogue
of the complex-$T$ ambiguity discussed above.}
In the
complex plane the final thermodynamic quantities depend on the way in which
the infinite volume limit is taken. From a physical point of view this is
not acceptable: we must obtain the same bulk thermodynamic quantities no
matter how we take the infinite volume limit. But again, this problem
fortunately only surfaces once we enter unphysical parameter values! The
need to take the thermodynamic limit in units of $4N$ instead of $N$ was
noticed already in ref. \cite{Shrock}, and in fact (as expected from the
example shown above), plays a highly non-trivial r\^{o}le in the proof. The
problem involved in taking the simplest infinite volume limit $N \to
\infty$ off the real axis becomes apparent if we consider the coupling
constant values $K + i(n+1/2)\pi$ discussed above. If we just let $N \to
\infty$, the partition function becomes a non-convergent oscillatory number
proportional to $e^{iN\pi/2}$.

The same remarks hold for other quantities, but not always with the same
restriction on only considering a particular thermodynamic limit.  For
example, although the correlation length $\xi = 1/\ln[\lambda_+/\lambda_-]$
does change under the transformation (\ref{eq:newSym2}), the {\em real part}
of its analytic continuation into the complex plane remains invariant.
Thus, Re($\xi^{-1}) = 0$ not only at the usual critical point(s), but also
at all points in the complex plane connected through the transformation
(\ref{eq:newSym2}). It was noted in ref. \cite{Shrock} that for the 2D Ising
model the condition Re$(\xi^{-1}) = 0$ is fulfilled on the same points in
the complex plane where the partition function vanishes. This is actually
no accident, occurring because $\xi \propto 1/\ln[\lambda_1/ \lambda_0]$,
where $\lambda_{1,0}$ are the two leading eigenvalues of the 2D Ising
transfer matrix \cite{Fisher2}. Since the 2D Ising partition function also
acquires its zeros when $|\lambda_0| = |\lambda_1|$, this is the rationale
for the equality of these two sets. This particular mechanism, degeneracy
of two (or more) leading transfer matrix eigenvalues, behind both
partition function zeros and the condition Re$(\xi^{-1}) = 0$, is in fact
more general \cite{Kac}. It occurs in the 1D Ising model as well. Here,
without a magnetic field, the condition Re$(\xi^{-1}) = 0$ implies
\beq
|\coth(K)| ~=~ 1,
\eeq
while the zeros of ${\cal Z}_N$ are found at
\beq
\lambda_+/\lambda_- ~=~ e^{i\pi(2n+1)/N},
\eeq
$i.e.$, for the $N \to \infty$ limit again given by the unit circle
$|\coth(K)| = 1$. In fact, it is precisely because it is also a partition
function zero that the condition Re$(\xi^{-1}) = 0$ in the complex plane
can be  associated with an unambiguous thermodynamic limit.

Similarly, the transformation (\ref{eq:newSym1}) also only defines a real
symmetry  of the 1D Ising model if a very particular definition of the
thermodynamic limit is used. The case $n=$  {\em even} is trivial, so we
need only consider the case $n=$ {\em odd}, $i.e.:~ K \to K + (2n+1)i\pi$.
Under this transformation, $ \lambda_{\pm} ~\to~ - \lambda_{\pm},$ and the
partition function is symmetric if one takes the thermodynamic limit in
units of $2N$. (Because the coordination number is $q = 2$, there is no
need to take the  thermodynamic limit in multiples of $4N$; this also
follows from the general proof in ref. \cite{Shrock}).

The existence of these kinds of symmetries appears to have been known for
some time by practitioners of transfer matrix techniques (see, $e.g$,
chapter 7 of ref. \cite{Baxter}).

\section{The Double-Scaling Limit: Matrix Models} \label{DoubleScal}

The non-analytic behaviour around phase transitions always arises as a result
of a singular procedure, the thermodynamic limit, in which the number of
degrees of freedom diverges. For a finite number of degrees of freedom, the
partition function is an analytic of the coupling (or temperature). It is
well known that a thermodynamic limit can be achieved not only by letting
the  volume (or number of sites) go to infinity, but also in finite volumes
(or even on just points, $i.e.$ zero dimensional field theories) provided
the number of degrees of freedom diverges by some other means.  One example
of this is large-$N$ matrix models in zero dimensions, which can display
behaviour completely analogous to ordinary infinite-volume phase
transitions. Another classical example is two-dimensional QCD in the  $N \to
\infty$ limit. By gauge invariance, two-dimensional  latticized $SU(N)$
Yang-Mills theory is a one-plaquette theory. This ``trivial" theory with
Wilson action nevertheless undergoes a 3rd order phase transition in the
limit  $N = \infty$ \cite{Witten}. One curious aspect of this 3rd order
Gross-Witten transition is that the correlation length, defined in a
natural physical way as
\beq
\xi \sim \sigma^{-1/2} , \label{eq:xi_UN}
\eeq
where $\sigma$ is the string tension, remains {\em finite} at the
phase transition. This is difficult to reconcile with renormalization
group ideas that seem to always require some diverging correlation
length at non-trivial fixed points. Does this mean that the
Gross-Witten phase transition cannot be understood from a renormalization
group perspective? Or do we just have to search a little more for the
true underlying correlation length that diverges at the transition?

The purpose of this section is therefore threefold.  First, we shall
analyze the matrix model phase transitions from the point of view of
scaling of partition function zeros. Second, we shall answer the question:
What is the analogue of the ``double scaling limit" \cite{Brezin}  of such
matrix models \cite{Ambjorn} in ordinary systems in statistical  mechanics?
And third, we shall describe the appropriate renormalization group scenario
for the Gross-Witten transition (which actually {\em does} correspond to a
diverging correlation length, to be defined shortly).

We start with the last two questions, which turn out to be intimately
related. But first a few definitions: By the double scaling limit of matrix
models (and low-dimensional string theory) we mean the following. The
singular part of the string partition function $\zeta(\Lambda,{\cal{G}})$
(a somewhat confusing terminology in this context, since it will be
identified with the {\em free energy} of certain matrix models) scales as
\cite{Polyakov}:
\beq
\zeta(\Lambda,{\cal{G}}) \sim (\Lambda^* - \Lambda)^{2-\gamma_0}
\sum_{G=0}^{\infty} a_G \left[(\Lambda^* - \Lambda)
{\cal{G}}^{-2/\gamma_1}\right]^G , \label{eq:stringPart}
\eeq
where the sum runs over genus $G$. Since the same factor appears for
all genera, this means that
\beq
\zeta(\Lambda,{\cal{G}}) \sim (\Lambda^* - \Lambda)^{2-\gamma_0}
F((\Lambda^* - \Lambda){\cal{G}}^{-2/\gamma_1}) \label{eq:stringPart2}~,
\eeq
where the scaling function $F$ is simply the sum on the r.h.s. of
eq.~(\ref{eq:stringPart}). Here $\Lambda$ is the cosmological constant, and
${\cal{G}}$ is the string coupling constant. The exponents $\gamma_0$ and
$\gamma_1$ are known from continuum calculations \cite{Polyakov}:
\begin{eqnarray}
\gamma_1 &=& \frac{1}{12}\left[25 - c + \sqrt{(1-c)(25-c)}\right] \cr
\gamma_0 + \gamma_1 &=& 2 , \label{eq:gammas}
\end{eqnarray}
where $c$ is the central charge. We shall mostly consider the
simple case $c=0$, which corresponds to zero-dimensional matrix models.

In such a matrix model representation, $(\Lambda^*-\Lambda) \sim (g^*-g)$
and ${\cal{G}} \sim 1/N$, where $g$ is the matrix model coupling constant,
and $N$ is the size of the matrix. The double scaling limit entails
sending $g \to g^*$, the critical point of the matrix model at $N = \infty$,
while simultaneously sending $N \to \infty$ in such a manner that
the scaling variable
\beq
x = (g - g^*)N^{2/\gamma_1}
\eeq
is kept fixed.

\begin{figure}[htb]
\begin{center}
\leavevmode
\epsfxsize=280pt
\epsfbox[170 350 520 510]{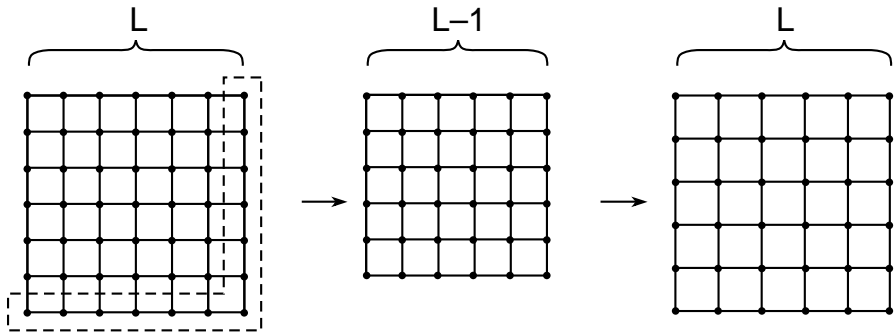}
\caption{\it Illustration of a real-space renormalization group
transformation.}
\label{fig:real_space_RG}
\end{center}
\end{figure}

What is the analogue of this double scaling limit in ordinary spin
systems? It turns out to be useful to think of a system at finite size $L$,
$i.e.$, volume $V=L^d$ in $d$ dimensions. A two-dimensional analogue of
such a system is shown in fig.~\ref{fig:real_space_RG}. Now let us perform
a simple real-space  renormalization group transformation in the form of a
spin decimation. Wishing to retain the same geometry (although in rescaled
form), we choose to integrate out all spins along one row and one column,
as also indicated in fig.~\ref{fig:real_space_RG}. Let us denote the spins
we integrate out by $\phi$, and the remaining ones by $\phi'$. This spin
decimation induces a renormalization group flow; the original Hamiltonian
$H$ of operators ${\cal{O}}_i$ and corresponding coupling constants $g_i$
is mapped into a new Hamiltonian $H'$ with in general new interactions
${\cal {O}}'_i$, and
in any case new coupling constants $g_i'$. That is,
\beq
{\rm e}^{L^dG(g)}\int d\phi' {\rm e}^{-H'(\phi',g')} = \int d\phi' d\phi
{\rm e}^{-H(\phi,\phi',g)}~,
\eeq
where we have explicitly, as is customary, extracted the constant part
$G(g)$ of the new Hamiltonian. Defining the free energy by
\beq
f(g) \equiv \frac{1}{L^d} \ln {\cal{Z}}(g) ,
\eeq
this indeed corresponds to the standard real-space
renormalization group equation for $f$,
\beq
f(g) = G(g) + b^{-d} f(g')
\eeq
with a length rescaling factor
\beq
b = \frac{L}{L-1} .
\eeq
But this time this equation has the meaning of a finite-size
renormalization relation. In the limit of large volume, $L \to \infty$,
the number of spins integrated out at each step compared with the
number of those retained becomes infinitesimal. This allows us to
phrase the real-space renormalization group equation in infinitesimal
form, giving in effect a Callan-Symanzik--type equation for
finite-size scaling.

The constant part $G(g)$ of the Hamiltonian is assumed to be regular. We
write it as
\beq
G(g) = \ln[\lambda(g)]
\eeq
where $\lambda(g)$ must approach unity for large $L$:
\beq
\lambda(g) = 1 + \frac{1}{L} r(g) +
{\cal{O}}\left(\frac{1}{L^2}\right) .
\eeq
Similarly, for large $L$, the coupling constants $g_i'$ will be
close to the previous values $g_i$, so that we may define functions
$\beta_i(g_i)$ by
\beq
g_i' = g_i + \frac{1}{L}\beta_i(g_i) +
{\cal{O}}\left(\frac{1}{L^2}\right) .
\eeq

It is then straightforward to see that the recursion relation
\beq
[\lambda(g)]^{L^d}{\cal{Z}}(L-1,g') = {\cal{Z}}(L,g)
\eeq
for the partition function implies the following differential
equation for the free energy $f(L,g)$:
\beq
\left[L\frac{\partial}{\partial L} - \sum_i \beta_i(g_i)\frac{\partial}
{\partial g_i} + d\right] f(L,g) =  r(L,g) . \label{eq:RoomWyld}
\eeq
This is the Callan-Symanzik equation for finite-size scaling. The
$\beta$-functions $\beta_i(g_i)$ have been used by
Roomany and Wyld \cite{Wyld} (see also ref. \cite{Barber}) to get
infinite-volume results from finite-size scaling. Their main advantage
is that they have zeros even in finite volumes, where no phase
transitions can occur. The zeros of these $\beta$-functions determine
the location of the phase transitions in the infinite-volume systems.

The Callan-Symanzik equation (\ref{eq:RoomWyld}) above gives an exact
differential equation for the free energy $f(L,g)$ for all values of the
couplings $g_i$ in the limit of infinite volume, $L \to \infty$. It is {\em
not} restricted to the critical regime close to a critical point given by
$\{g_i^*\}$.

For large but finite volumes $L^d$, we can extract the singular behaviour
of the free energy near the infinite-volume fixed point $\{g_i^*\}$. For
simplicity, assume that we have only one coupling constant $g$.
Define an ordinary critical point $g^*$ by a solution of
\beq
\beta(g^*) = 0~,~~~~~ \beta'(g^*) > 0 .
\eeq
Then the Callan-Symanzik equation (\ref{eq:RoomWyld}) gives the following
solution for the singular part of the free energy $f(L,g)$ near $g^*$:
\beq
f(L,g) \sim (g^* - g)^{d/\beta'(g^*)} F((g^*-g)L^{\beta'(g^*)})~,
\label{eq:freeE_FSS}
\eeq
where $F$ is an arbitrary scaling function. This of course coincides
with the usual finite-size scaling formula if we identify the
correlation length exponent as follows:
\beq
\nu = \frac{1}{\beta'(g^*)} ,
\eeq
an identification that can also be made explicitly by considering the
rescaling of the correlation length under the spin decimation
described above. Note that the regular part $r(L,g)$ of the
Callan-Symanzik equation (\ref{eq:RoomWyld})
does not influence the singular part of the free energy, as expected.

While this formulation thus allows us to recover the standard
finite-size scaling formula, it actually also gives us more. It
relates the divergence of the correlation length directly to the
rescaling of couplings through their $\beta$-functions. This will
turn out to be useful when we return to matrix models.

Implicit in the above considerations was the assumption of an analytic
expansion of the $\beta$-function around $g^*$,
\beq
\beta(g) = \beta(g^*) - \beta'(g^*)(g^*-g) + {\cal{O}}\left(
\frac{1}{L^2}\right)
\eeq
and we also had to assume that $\beta'(g^*) > 0$, since otherwise
we would not arrive at a positive correlation length exponent.
A negative correlation length exponent would be inconsistent with
the assumption of $g^*$ being a critical point. More exotic
phase transitions are, however, also contained in this formalism.
Such phase transitions correspond to more unusual behaviour of the
$\beta$-functions \cite{Wyld}.

Note that the finite-size scaling form of the free energy $f(L,g)$ has an
uncanny resemblance to the Knizhnik-Polyakov-Zamolodchikov scaling form of
the string partition function (\ref{eq:stringPart2}). For this
identification to be made, we see from the second relation of
eq.~(\ref{eq:gammas}) that we have to consider the string (matrix model)
case as corresponding to $d=2$, and identify the size $N$ of the matrix
with the size $L$ of the system described above. This is actually entirely
natural, since as far as scaling in size $N$ is concerned, a matrix acts
like a 2-dimensional system.  The precise identification is then
\beq
\gamma_0 = 2 - \frac{2}{\beta'(g^*)}~,~~~~~~\gamma_1 = \frac{2}{\beta'(g^*)}~.
\label{eq:gam_betP}
\eeq
To conclude, {\em
the double scaling of matrix models is precisely the
finite-size scaling} of ordinary 2-dimensional statistical mechanics.

In matrix model language, the finite-size scaling -- as we have seen,
the ``double scaling" -- arises in the limit of $N \to \infty$, when
one simultaneously tunes the matrix coupling $g$ to a critical value
$g^*$ in such a manner that the scaling variable
\beq
x \equiv (g^*-g)N^{2/\gamma_1}
\eeq
is kept fixed. The derivation of finite-size scaling presented here for
spin systems was first done for matrix models by Carlson \cite{Carlson}
through the entirely analogous procedure of integrating out one row and one
column of an $N\times N$ model, thereby relating it to an
$(N-1)\times(N-1)$ model with renormalized coupling constants. It has very
recently been rediscovered by Br\'{e}zin and Zinn-Justin \cite{Brezin1},
who noted the identification (\ref{eq:gam_betP}) above
(see also ref.\cite{Alfaro}).
In fact, in order to facilitate the comparison with the
work of Br\'{e}zin and Zinn-Justin  \cite{Brezin1}, we have used a notation
that closely parallels  theirs in the language of matrix models.

This interpretation of the double-scaling limit of matrix models
implies immediately an almost trivial relation between the string
susceptibility $\gamma_0$ (and, equivalently, $\gamma_1$) and
the correlation length exponent $\nu$:
\beq
\nu =  \frac{1}{2}(2-\gamma_0) = \gamma_1/2 . \label{eq:gam_nu}
\eeq

One obvious generalization of these relations is to rank-$d$ tensor models
(where ``tensor" is just a loose terminology for any $d$-index object).
Vector models have already been analysed from the  double-scaling point of
view \cite{Ambjorn1}, and found to behave very similarly to the matrix
models. Tensor models of higher rank tensors have also been suggested in
relation to simplicial gravity in higher dimensions \cite{Gross}. We see
from the straightforward scaling arguments above that for simple
zero-dimensional tensor models, the double scaling limit is uniquely given
by the rank of the tensor. It is natural to parametrize the critical
exponents of rank-$d$ tensor models slightly differently from the matrix
model case, in order to avoid factors of $d$ in some of the expressions.
Let us therefore write the free energy for rank-$d$ tensor models in the
form
\beq
f(N,g) \sim (g^*-g)^{2-\gamma_0} F((g^*-g)N^{d/\gamma_1}) ~.
\eeq
This, then, defines the two exponents $\gamma_0$ and $\gamma_1$.
Comparing with eq.~(\ref{eq:freeE_FSS}), we note that with this definition,
the relation
\beq
2 - \gamma_0 = \gamma_1
\eeq
remains valid for all rank-$d$ tensor models. The correlation length
exponent $\nu$ is then related in a very simple manner to the string
susceptibility exponents $\gamma_0$ and $\gamma_1$:
\beq
\nu = (2 - \gamma_0)/d = \gamma_1/d ~. \label{eq:gam_nu_d}
\eeq

The correlation length has a very direct interpretation in terms of the
rank-$d$ tensor models. It describes the range of effective correlations
within invariant subgroups of the large-$N$ tensor, and it diverges at the
critical point. Without this diverging correlation length, the description
of the double scaling limit in renormalization group terms would presumably
be  void of meaning.

It is interesting to compare these simple scaling laws with what has
already been derived by other means. We have earlier seen the direct
relation between finite-size scaling of matrices and the
Knizhnik-Polyakov-Zamolodchikov formula for strings. Another class of
models which have been studied in the double-scaling limit is that of
vector models \cite{Ambjorn1}, which in our notation corresponds to $d$ =
1. The critical exponents for these vector models are precisely given by
the relations (\ref{eq:gam_betP}) and (\ref{eq:gam_nu_d}) with $d = 1$.

The interpretation of the correlation length in terms of the original
``dual" picture (random polymers ($d$ = 1), random surfaces ($d$ = 2),
etc.) is unfortunately still rather obscure.  The divergent correlations
occur {\em within the rank-d tensor itself}, and does not seem to have any
simple physical meaning. Similarly, if we return to the solvable case of
2-dimensional lattice-regularized Yang-Mills theory, there is no
contradiction between the finite physical correlation length
(\ref{eq:xi_UN}) which is defined in terms of the lattice correlator, and
the divergent correlation length occurring among subsets of matrix elements
inside the $SU(N)$ matrix as $N \to \infty$. The latter has no simple
interpretation in terms of the usual Yang-Mills potentials.

Nevertheless, having a well-defined diverging correlation length available
makes it possible for us to make use of the full machinery of
the theory of complex-plane
partition function zeros discussed in this paper. This in particular
implies that also the double-scaling limit itself can be entirely
understood from the point of view of  singularities in the complex coupling
constant plane. This is the subject of the next subsection.

\subsection{\sc{Two-dimensional QCD in the large-$N$ limit}}

As an application of the above ideas, we shall now demonstrate how the
scaling of complex coupling constant singularities can be used to
numerically determine, $e.g.$, the string susceptibility exponents. The
single crucial ingredient is the existence of the relation
(\ref{eq:gam_nu}) between the correlation length exponent $\nu$ and the
matrix model susceptibility exponent $\gamma_0$.

For numerical purposes it turns out that the unitary matrix model --
2-dimensional latticized Yang-Mills theory -- is the easiest matrix model
to work with. The relation between finite-size scaling and the
double-scaling limit as discussed above was derived only in the case of a
Hermitian matrix ensemble, where the integration over a row and a column
does not spoil hermiticity. In the unitary ensemble, the mapping onto
itself no longer holds.  For finite values of $N$, the size of the group
elements in the fundamental representation, one is not automatically left
with a unitary matrix after integrating out a row and a column. On the
surface, this would seem to  invalidate the direct connection with
finite-size scaling. Indeed, an exact integration of a row and column can
only yield a unitary matrix model modulo $CP^{N-1}$ factors \cite{Carlson}.
Fortunately, these corrections are subleading in the large-$N$ limit, and
the above formalism therefore {\em does} carry through in the limit $N \to
\infty$.

As we are interested in the large-$N$ limit, where the distinction between
$SU(N)$ and $U(N)$ is immaterial, we restrict ourselves to 2D lattice
gauge theories of the $U(N)$ kind. With the Wilson action one can write
the reduced partition function as
\beq
{\cal{Z}}_N ~=~ \int [dU_P] e^{g^{-2}[Tr U_P + Tr U_P^{\dagger}]} ,
\label{eq:UN}
\eeq
where variables have already been changed to fundamental plaquettes $U_P$,
and where a trivial volume factor has been taken out.  This partition
function can be written in closed form for any $N$ by means of modified
Bessel functions \cite{Bars}:
\beq
{\cal{Z}}_N ~=~ \det(M_{ij}) ~=~ \det(I_{i-j}(2/g^2))~,~~~~~i,j ~=~
1, 2, \ldots, N~, \label{eq:BessDet}
\eeq
and it is this expression we shall use to determine the partition function
zeros. In order to be able to take a smooth large-$N$ limit, we rescale
variables and introduce $\lambda = g^2N$. The Gross-Witten phase transition
occurs at $\lambda^* = 2$ \cite{Witten}.

The critical exponent $\gamma_1$ from the double-scaling limit of this
unitary matrix model was computed soon after the discovery of the
double-scaling limit in the Hermitian matrix ensemble \cite{Periwal}. In
the notation of this  paper, the result is $\gamma_1$ = 3. This
corresponds, according to eq.~(\ref{eq:gam_nu}), to a correlation length
exponent
\beq
\nu = 3/2 ~. \label{eq:nu_UN}
\eeq

Next, let us consider the model (\ref{eq:UN}) in the complex
$\lambda$-plane. According to the general scaling theory of complex-plane
zeros of the partition function \cite{Itzykson}, the partition function
zeros closest to the Gross-Witten transition point $\lambda^* = 2$ approach
this transition point at a rate determined by the  correlation length
exponent $\nu$:\footnote{At the upper critical dimension, this behaviour is
modified by logarithmic corrections \cite{Lang}.}
\beq
{\mbox{\rm Im}} (1/\lambda_0) \propto N^{-1/\nu}~.   \label{eq:FSS_nu2}
\eeq

One can easily show \cite{Kolbig} that
\beq
{\cal{Z}}_N (\lambda) = \overline{ {\cal{Z}}_N (\bar \lambda)} =
{\cal{Z}}_N (-\lambda) .
\eeq
Thus each zero of ${\cal{Z}}_N (\lambda)$ comes in quadruplets, $\lambda_0$,
$-\lambda_0$, $\bar \lambda_0$, and $- \bar \lambda_0$, unless it is purely
imaginary. We have computed the zeros $1/\lambda$ with the smallest
imaginary part for $N = 1$ to 10. Most of these are already listed in
\cite{Kolbig}, but for completeness we give the representative in the first
quadrant in table~\ref{tab:Bess_zeros}.\footnote{The exact analytical form
of the domain of partition function zeros in the $N = \infty$ theory can
in principle be calculated analytically using the techniques of
ref.\cite{David}. This is an interesting open problem which remains to
be solved. The approximate form of the partition function zeros in the
$N = \infty$ theory has been conjectured in ref.\cite{Kolbig} on the
basis of numerical evidence for finite values of $N$.}

\begin{table}[hbt]
\begin{center}
\caption{\it The partition function zeros of eq.~(\protect \ref{eq:UN})
closest to the real axis.}
\label{tab:Bess_zeros}
\vskip 0.5cm
\begin{tabular}{|ccc|}
\hline
 $N$ & Re($2/\lambda_0$) & Im($2/\lambda_0$) \\
\hline
  1 &   0.0      &  2.404825 \\
  2 &   0.639801 &  1.490191 \\
  3 &   0.810578 &  1.130558 \\
  4 &   0.883926 &  0.930444 \\
  5 &   0.922905 &  0.800366 \\
  6 &   0.946328 &  0.707894 \\
  7 &   0.961580 &  0.638198 \\
  8 &   0.972091 &  0.583456 \\
  9 &   0.979646 &  0.539118 \\
 10 &   0.985256 &  0.502343 \\
\hline
\end{tabular}
\end{center}
\end{table}

\begin{figure}[htb]
\begin{center}
\leavevmode
\epsfxsize=280pt
\epsfbox[20 30 620 500]{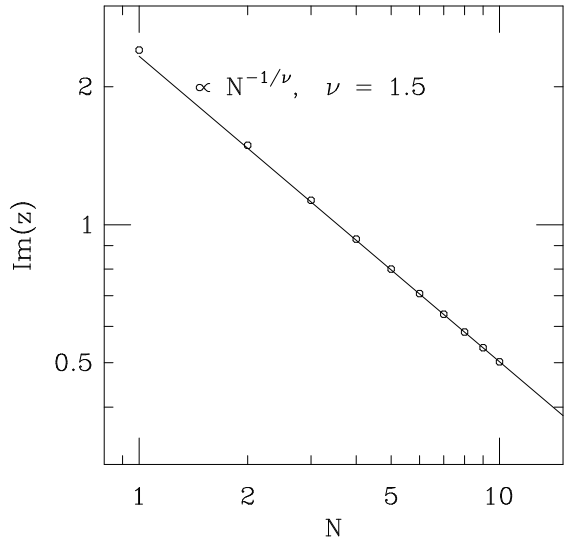}
\caption{\it ``Finite $N$" scaling of the complex partition function zero
closest to the real axis for the Gross-Witten model.}
\label{fig:Zero_scal_UN}
\end{center}
\end{figure}

One can now use the finite size, {\it i.e.,} in our case finite $N$,
scaling relation (\ref{eq:FSS_nu2}) to obtain estimates of $\nu$. Such
estimates, obtained from $N$ and $N-1$ are given in
table~\ref{tab:nu_zeros_UN}. These estimates still have subleading finite $N$
corrections. Assuming a simple $1/N$ behaviour we can obtain, from two
successive $\nu(N)$'s estimates for the true, infinite $N$ values:
\beq
\nu(\infty) = \nu(N) + (N-1)[\nu(N) - \nu(N-1)] ~.
\eeq
These estimates for $\nu(\infty)$ are also listed in
table~\ref{tab:nu_zeros_UN}. As one can see, the convergence to the
theoretical value 3/2, eq.~(\ref{eq:nu_UN}), is very fast. This can also
be seen from fig.~\ref{fig:Zero_scal_UN}.

\begin{table}[hbt]
\begin{center}
\caption{\it Estimates of $\nu$ from the ``finite $N$" scaling of the
imaginary part of the partition function zeros computing from successive
$N$ (second column) and extrapolated to $N = \infty$ from two successive
estimates (third column).}
\vskip 0.5cm
\label{tab:nu_zeros_UN}
\begin{tabular}{|ccc|}
\hline
 $N$ & $\nu(N)$ & $\nu(\infty)$ \\
\hline
  2 &  1.448363 &  ~~ \\
  3 &  1.468050 &  1.507424 \\
  4 &  1.476775 &  1.502950 \\
  5 &  1.481765 &  1.501725 \\
  6 &  1.485009 &  1.501227 \\
  7 &  1.487288 &  1.500966 \\
  8 &  1.488970 &  1.500745 \\
  9 &  1.490265 &  1.500620 \\
 10 &  1.491294 &  1.500561 \\
\hline
\end{tabular}
\end{center}
\end{table}

{}From the exact solution by \cite{Witten} we can also see that $\alpha =
-1$. Using $d = 2$ as appropriate for the matrix model considered, we can
see that $\alpha$ and $\nu$ satisfy the usual (hyper-) scaling relation,
eq.~(\ref{eq:hyperscal}). If this is not fortuitous, and we see no
reason why this should be the case, it implies that the string
susceptibility exponent $\gamma_1$ of the double-scaling limit could have
been obtained directly from the Gross-Witten solution, using the above
identifications.

An interesting question concerns the possible existence of an order
parameter describing the phase transition. With the correct notion of
a magnetic field operator, one could establish the missing critical
index, and derive the corresponding scaling relations. It has been
suggested \cite{Celmaster} that the Gross-Witten phase transition can
be associated with the symmetry breaking of finite $U(N)$ subgroups of
the full $N \to \infty$ theory. However, the ``symmetry breaking" of
ref. \cite{Celmaster} occurs on {\em both} sides of the phase transition
point, and it is not immediately obvious how it connects to the
conventional notion of an order parameter. This aspect of the $N =
\infty$ phase transition remains to be better understood.

\section{\sc{Conclusions}} \label{Conclusions}

The generalization of spin, matrix and gauge models to complex coupling
constants yields new and useful insight into the dynamics of these
systems. We have discussed a selection of solvable or near-solvable
models where predictions of the complex-plane behaviour can be tested
directly. The one-dimensional Ising and generalized Heisenberg models
are particularly useful examples of theories with non-trivial phase
transitions (albeit, in the Ising model case, at zero temperature)
where one can study the behaviour in both the complex activity and
temperature planes. By exact decimation, one can also describe the
behavior of renormalization group trajectories in these complex planes.

We have discussed some of the subtleties involved in classifying new
symmetries of such spin models in the complex temperature plane. The
question of whether these models have new symmetries hinges on the
precise definition of what is meant by the thermodynamic limit for
theories with complex Boltzmann factors or transfer-matrix eigenvalues.

Matrix models have recently been under intense study in connection with
the ``double-scaling" limit of low-dimensional string theory. We have
clarified various aspects of this double-scaling limit by showing that
it is equivalent to what in standard statistical mechanics terminology
is known as finite-size scaling. Here ``size" ($i.e.$ volume) is
nothing but the size $N$ of an $N\times N$ matrix. The double scaling
of matrix theory is thus equivalent to the finite-size scaling of
{\em two}-dimensional systems, while the corresponding double scalings
of rank-$d$ tensor models are equivalent to finite size scalings of
$d$-dimensional systems in statistical mechanics. This remarkable
connection allows us to phrase the string susceptibility exponent
of matrix models directly in terms of the correlation length exponent
$\nu$. By entering the complex coupling constant plane, the scaling
of partition function zeros near the critical point is thus given
in terms of the string susceptibility exponent. Conversely, we have
demonstrated how this scaling of partition function zeros can be
used to give remarkably accurate values for the string susceptibility
exponent. As a by-product, we have established that the 3rd order
Gross-Witten phase transition of 2-dimensional $U(N)$ ($N \to \infty$)
lattice gauge theory with Wilson action is characterized by an
underlying divergent correlation length, despite the finite value of
the naive
correlation length \cite{Witten}. The Gross-Witten phase transition
of 2-dimensional Yang-Mills theory at large $N$ can thus be understood
in entirely conventional renormalization group terms.

\end{document}